\documentclass[12pt]{article}
\usepackage{graphicx,adjustbox,amsmath,amssymb,lineno,url,authblk,setspace,ulem}

\newcommand{\avg}[1]{\left \langle #1  \right\rangle}

\begin{document}
\def\correspondingauthor{\footnote{achacoma@famaf.unc.edu.ar}}

\title{Simple mechanism rules the dynamics of volleyball}
\author{Andr\'es Chacoma\correspondingauthor{} }
\author{Orlando V. Billoni}

\affil{Instituto de F\'isica Enrique Gaviola (IFEG-CONICET), Ciudad Universitaria, 5000 C\'ordoba, Argentina}
\affil{Facultad de Matem\'atica, Astronom\'ia, F\'isica y Computaci\'on, Universidad Nacional de C\'ordoba, Ciudad Universitaria, 5000 C\'ordoba, Argentina}

\date{}
\maketitle

\begin{abstract}
In volleyball games, we define a rally as the succession of events observed since the ball is served until one of the two teams on the court scores the point.
In this process, athletes evolve in response to physical and information constraints, spanning several spatiotemporal scales and interplaying co-adaptively with the environment. 
Aiming to study the emergence of complexity in this system, we carried out a study focused on three steps: data collection, data analysis, and modeling.
First, we collected data from 20 high-level professional volleyball games. 
Then we conducted a data-driven analysis from where we identified fundamental insights that we used to define a parsimonious stochastic model for the dynamics of the game. 
On these bases, we show that it is possible to give a  closed-form expression for the probability that the players perform $n$ hits in a rally using only two stochastic variables. 
Our results fully agree with the empirical observations and represent a new advance in the comprehension of team-sports competition complexity and dynamics.
\end{abstract}

\section{Introduction}

The engaging nature of the complexity related to team sport competition has historically captured the attention of academics 
\cite{blasius2009zipf,clauset2015safe,holleczek2012particle,baek2015nash,ben2007efficiency,yamamoto2021preferential,neiman2011reinforcement,mukherjee2019prior,mandic2019trends,merritt2013environmental,gudmundsson2017spatio,fister2015computational}.
In recent years, the interest in studying this area has increased. Promoted by the demands of the thriving global sports market, the research community is now seeing the study of the phenomena related to these systems as challenges to face in the context of a hyper-competitive twenty-one-century industry \cite{patel2020intertwine}.
However, most of the papers available in the literature use empirical data to perform statistical analysis aiming to generate key performance indicators \cite{drikos2009correlates,buldu2019defining,bransen2019measuring,cakmak2018computational,gama2016networks,garrido2020consistency}. 
This approach can be handy for coaches using data-driven based training systems, but it usually does not provide a deeper understanding of the phenomena.
From an alternative perspective, our view focuses on use empirical data to uncovering the mechanisms that explain the emergence of global observables;
based on a multidisciplinary framework that connects physics with complexity in the social sciences \cite{jusup2022social}.
In previous works, we have successfully studied an modeled several aspects of  the game of football \cite{chacoma2020modeling,chacoma2021stochastic}. 
In this paper, we report a strikingly simple mechanism that governs the dynamics of volleyball. 

To place the complexity of this sport in context, let us briefly discuss the basic concepts.
In volleyball games, two teams of six players, separated by a high net, try to score points by grounding the ball on the other's team court.
In possession of the ball, teams may hit the ball up to three times, but players must not hit the ball twice consecutively.
During a rally, namely the time between the serve and the end of the play, the teams compete for the point using specially trained hits to control the ball, set, and attack.  
Blocks, likewise, are trained actions to try to stop or alter an opponent's attack, but unlike the others, the hit involved in blocks does not count as an ``official'' hit. It means that a player who hits the ball by blocking can consecutively touch the ball again without infringing the rules.
On the other hand, we want to point out that the emergence of complexity in the game is strictly related to players' ball control and accuracy. 
Note that if one of the players is not able to handle the ball, their team misses the point; if one of the players cannot pass the ball accurately, it increases the probability that the teammates can not control the ball, and if one of the players cannot attack accurately, it increases the probability that the opponents can handle the ball and eventually obtain the point.

In order to study this complex dynamic and unveil the mechanisms behind it, we proceed as follows. 
First we visualized and collected data from 20 high-level international games.
Then we performed a data-driven analysis to get insight into the underlying mechanisms of the game. 
With this, we proposed a parsimonious model and developed an analytical approach to obtain a closed-form expression able to capture in remarkable good approximation one of the most relevant observables of the dynamics: the probability that the players perform n hits in a rally, $P_n$.
A variable that is key to characterizing the system complexity, and it is also related to teams performances \cite{sanchez2015analysis,link2019performance}, it is involved in self-regulated phenomena \cite{sanchez2016dynamics}, may affect the motor activities during a top-level match \cite{mroczek2014analysis}, etc.

\section{Data collection}

To collect the data used in this work, we coded a visualization program that allows extracting information from video records of volleyball games.
The program is designed for a trained operator to record the most relevant events observed during rallies. The information gathered at each event  include,
\begin{enumerate}
\item An identification number for the player that hit the ball in the event.

\item The time of the event referred to the beginning of the game.

\item The position of the player that hit the ball. In two dimensions, referred to the court's floor.

\item The type of hit performed: pass, set, attack, block, etc.
\end{enumerate}

We visualized $10$ games of the 2018 FIVB Men's Volleyball Nations League and $10$ games of the 2019 FIVB Men's Volleyball Challenger Cup, collecting the information of $3302$ rallies in total. 
The visualization program and an anonymized version of the collected data used for this work can be found in \cite{data}. For further information on the visualized games and details of the visualization process, please c.f. Supplementary Material section S1.

\section{Data analysis}

\begin{figure}[t!]
\centering
\includegraphics[width=0.75\textwidth]{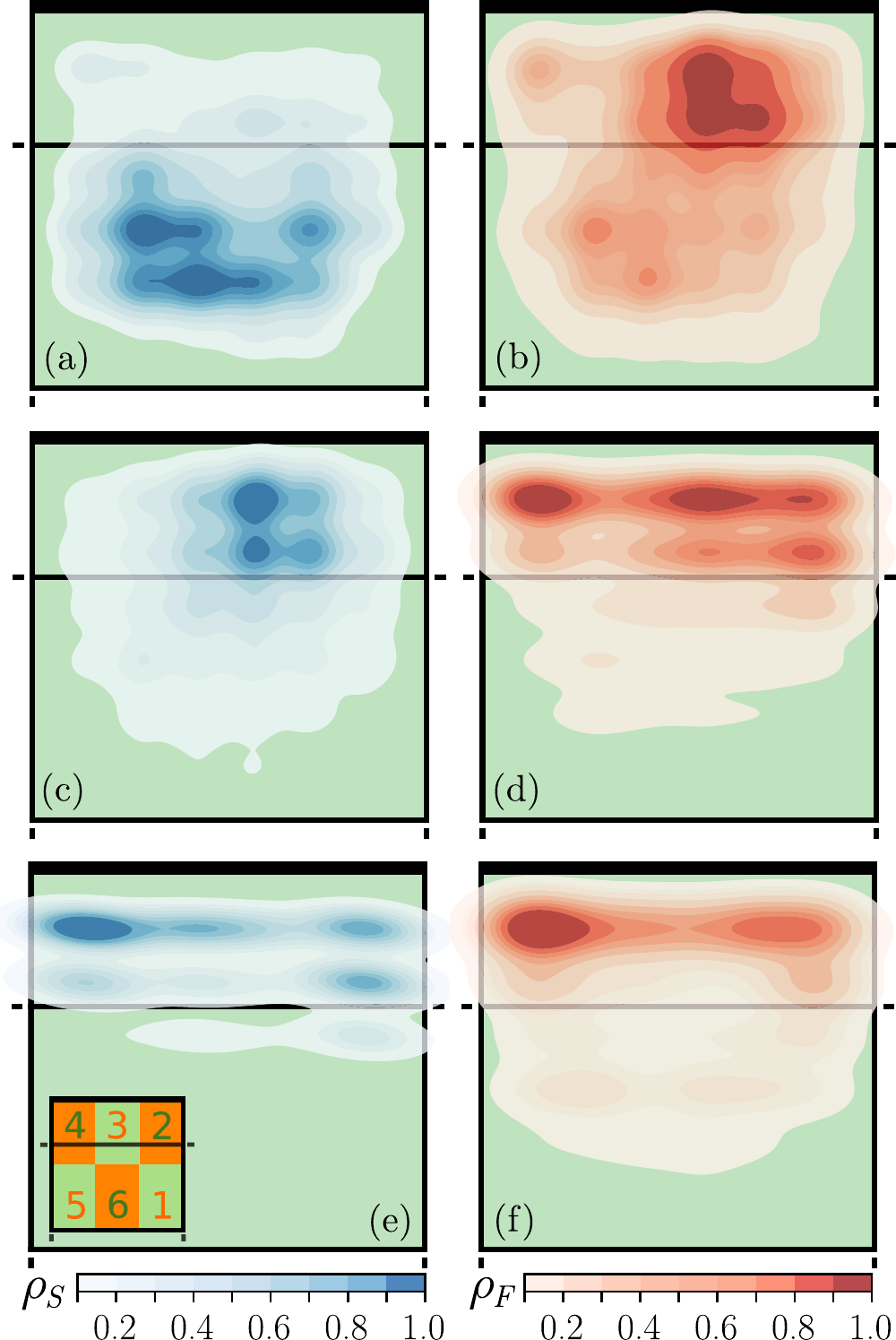}
\caption{
Probability that the players hit the ball in particular zones of the court.
Notice that in each panel, we show the court layout, including the position of the net at the top, the 3 meters line at the middle, the background line at the bottom, and the lateral lines at the sides.
Panels (a)-(b) show the case of the 1st hit, (c)-(d) of the 2nd hit, and (e)-(f) the case of the 3rd hit.
Plots at the left (colored in blue) shows the results when the possession succeeds.
Plots at the right (colored in red) when they miss the point.
The inset in panel (e) shows the six zones of the court used as reference.
We use the information presented in these plots to highlight the lack of performance that the players exhibit when moving out of their action zones to perform a hit.
}
\label{fi:insight}
\end{figure}

We analyzed the collected data to understand why the players succeed or fail when they intend to score a point.  
In Fig.~\ref{fi:insight} we show heatmaps indicating the probability that the players hit the ball in particular zones of the court. Panels (a)-(b) show the case of the 1st hit, (c)-(d) of the 2nd hit, and (e)-(f) the case of the 3rd hit.
At the left (colored in blue), we show the results when the possession succeeds, namely when they scores the point. At the right (colored in red), when they miss the point.
Parameters $\rho_S$ and $\rho_F$ give the probability normalized by the maximum value and are linked to the color intensity of the heatmaps.
In the inset of panel (e), we show the court divided and numerated into zones. Notice that, in the following, we will use these references to discuss the results.

As a first observation, we can see that, in successful possessions, it is highly probable that the players hit the ball in their natural action zones: the first hit (reception) in zones 5-6-1, the second one (set up) around 2-3, and the attack in the zones 4-3-2 (at the front) and 6-1 (at the back). 
In unsuccessful possessions, we can see that the players have to move out of the actions zones to perform the hit. 
For instance, in panel (b), we can see that the probability of the players performing the 1st hit around zone 2 rises. 
Note, that it is not tactical convenient because this is the zone that the setter uses to perform the 2nd hit, and the tight presence of other players may hinder them and produces a fail or cause a reduction of accuracy.
If the 1st hit is performed inaccurately, it produces what we see in panel (d), where the setter frequently has to go to the attackers' zones to handle the ball causing the same hinder effect as in the previous case. 
In this context, the attackers' options are limited. 
In extremes cases, they will not attack. They will pass an easy ball to the opponents just to keep playing.
Therefore, in these cases, the players' performance diminishes. 
In conclusion, when the players have to move out of their action zones to handle the ball, it increases the probability of both missing the ball and hitting the ball imprecise.
In the following, we will use these observations to define our model.

\section{Model}

We have observed that, during rallies, when the players have to move out of their action zone to perform a hit, the probability of missing the ball increases, and the precision decreases.
In the light of these observations, to model the rallies' dynamic, we introduce two stochastic parameters, $p$ and $q$, as follows,
\begin{enumerate}
    
    \item {\it The probability of performing the hit, $p$}.
    If the players have to move out of their action zones to hit the ball, then, there is a probability $p$ of performing the hit, and $1-p$ of missing it. 
    
    \item {\it The probability of achieving precision, $q$}. 
    If the players have to move out of their action zones to perform a hit, then there is a probability $q$ of achieving precision in two situations: 
    \begin{enumerate}
        \item Defending or setting, in the first and second hit passing the ball towards the teammate's action zone.
        
        \item Attacking or serving, sending the ball out of the action zone of the opponent in charge of taking the first hit. 
    \end{enumerate}
    
    With probability $1-q$, likewise, the opposite occurs in both situations.
\end{enumerate}
Following these rules, the players will move the ball around the court $n$ times until one of them misses a hit, ending the rally.

\section{Analytical approach to obtain $P_n$}
\label{se:analytical}

We now focus on developing an analytical approach to obtain the probability distribution that the players perform $n$ hits during rallies, $P_n$.
First, we introduce two approximation: (i) we will suppose that the teams always try to use the three reglementary hits to score the point. In other words, they do not pass the ball to the other team at the first or second hit. Considering that these events are rare in the dataset ($<0.1\%$ of the observed cases), we understand that this is a reasonable approximation; and
(ii) we will not consider the blocks as hits. By rule, when a team blocks an attack, they still have the three hits to control, set and attack. Therefore, in practice, the effects of blocks can be absorbed as inefficiencies in the attack. 

Let us start calculating the cases $P_1$, $P_2$ and $P_3$. We obtain,
\begin{subequations}
\begin{align}
P_1 &= q\,(1-p), \label{eq:P1}\\
P_2 &= q\,p\,(1-q)\,(1-p), \label{eq:P2}\\
P_3 &= q\,p\,(1-q)\,p\,(1-q)\,(1-p). \label{eq:P3}
\end{align}
\label{eq:1}
\end{subequations}
In eq.~(\ref{eq:P1}), the probability $q$ indicates that the service is performed outside the action zone of the player in charge of taking the first hit (a difficult service), and the probability $(1-p)$ indicates that this player cannot perform the hit. Consequently, in the rally only one hit is performed: the service.
In eq.~(\ref{eq:P2}), $q$ indicates a difficult service, $p$ indicates that the player can perform the second hit, $(1-q)$ indicates the pass is performed outside the action zone of the player in charge of taking the second hit (setter), and $(1-p)$ indicates that the later cannot perform the third hit. Consequently, in the rally two hits are performed: the service, and the first hit.
With a similar analysis, we can obtain the probability of performing three hits, $P_3$, that we show in eq.~(\ref{eq:P3}).

\begin{figure}[t!]
\centering
\includegraphics[width=0.8\textwidth]{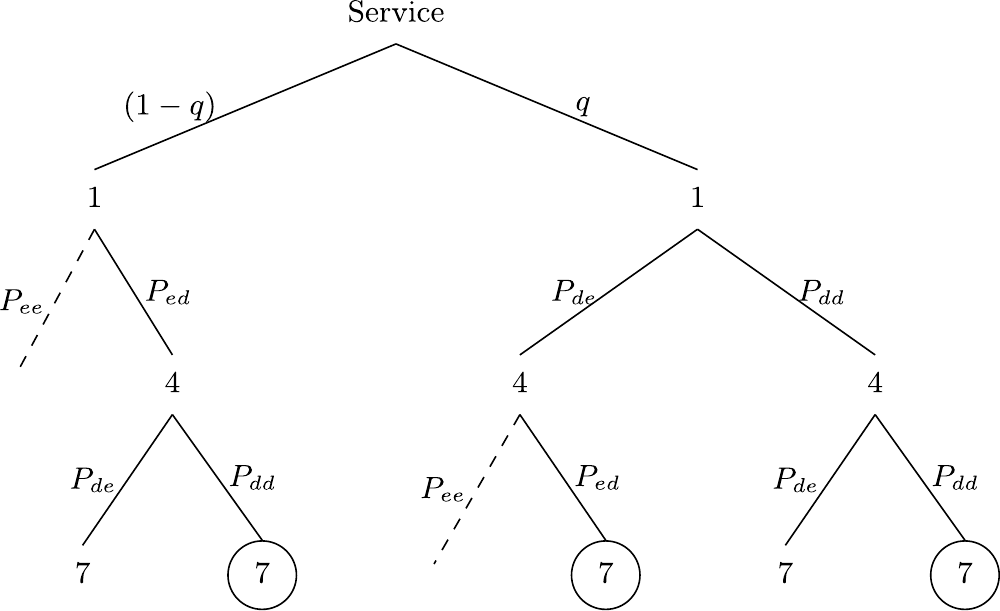}
\caption{
With transition probabilities $P_{dd}$ , $P_{de}$, $P_{ed}$, and $P_{ee}$, it is possible to evaluate the probability that  the players perform $n$ hits in a rally, $P_n$, by using a decision tree. In this plot we show the
decision tree used to calculate $P_7$. 
Nodes represent the number of hits performed in complete ball possession and edges transition probabilities. 
Circles show the end of the paths used for the calculation of probability $D_3$ [see main text, Eq.~(\ref{eq:4})].
} 
\label{fi:Decisiontree}
\end{figure}

To obtain $P_4$ and further values of $n$ is useful to define the transition probabilities linked to the teams' chances of performing the three hits that define a complete possession,  
\begin{equation}
\begin{split}
\centering
P_{dd} &= p\,q+ p\,(1-q)\,p\,q + p\,(1-q)\,p\,(1-q)\,p\,q, \\
P_{de} &= p\,(1-q)\,p\,(1-q)\,p\,(1-q),\\
P_{ed} &= 1, \\
P_{ee} &= 0,
\end{split}
\label{eq:2}
\end{equation}
where $P_{dd}$ is the probability that a team receives a difficult ball and, after the three hits, delivers to the other team a difficult one, $P_{de}$ is the probability that a team receives a difficult ball and delivers an easy one, $P_{ed}$ the probability that a team receives an easy ball and delivers a difficult one, and $P_{ee}$ the probability that a team receives and delivers an easy ball.
In this frame, $P_4$ can be written as follows,
\begin{equation}
\centering
P_4 = q\,P_{dd}\,(1-p) + (1-q)\, P_{ed}\,(1-p),
\label{eq:3}
\end{equation}
where the first term is the probability that a team receives a difficult ball, handles it, delivers a difficult ball, and the other team cannot achieve the first hit. The second term is the probability that a team receives an easy ball, delivers a difficult one, and the opponent team cannot achieve the first hit.

To calculate $P_n$ for higher values of $n$, we use a binary decision tree. 
As an example, let us focus on calculating $P_7$. 
In Fig.~\ref{fi:Decisiontree} we exhibit a three levels tree where the nodes indicate the number of hits performed and the edges the transition probabilities to the next level.  
Note, the level of the tree, $l$, is related to the number of performed hits by the expression $n=3l-2$. 
If we define the probability $D_3$ as,
\begin{equation}
\centering
D_3 = 
q\,P_{dd} P_{dd} +
q\,P_{de} P_{ed} +
(1-q)\,P_{ed} P_{dd},
\label{eq:4}
\end{equation}
obtained from the sum of all the paths leading to the third level leaves indicating that the team in possession is delivering a difficult ball (see circles in Fig.~\ref{fi:Decisiontree}),
then, to obtain $P_7$ we just multiply for $(1-p)$, which indicates the team receiving the attack cannot perform the $8th$ hit. Therefore,
\begin{equation}
\centering
P_7 = D_3\,(1-p).
\label{eq:5}
\end{equation}

To calculate the probability for values of $n$ between two levels of the tree, for instance $P_8$ and $P_9$, we use Eq.~(\ref{eq:4}) as follows,
\begin{equation}
\centering
\begin{split}
P_8 &= D_3\,p\,(1-q)\,(1-p)\\
P_9 &= D_3\,p\,(1-q)\,p\,(1-q)\,(1-p).
\end{split}
\label{eq:6}
\end{equation}

As the reader may be noted, the calculation of $D_l$  is useful to obtain $P_n$. Therefore in the following, we focus on calculating $D_l$ $\forall \, l\in \mathbb{N}$.
We can write the probability of ``achieving'' a level $l$ of the tree, $A_l$, as,
\begin{equation}
\centering
A_l = E_l + D_l,
\label{eq:7}
\end{equation}
where probability $E_l$ is the sum of all the paths leading to the $l\,th$ level leaves which transition probabilities indicate that the team in possession is delivering an easy ball.
Similarly, we can write $D_{l+1}$ and $A_{l+1}$ as,
\begin{equation}
\begin{split}
\centering
D_{l+1} &= E_l\,P_{ed} + D_l\,P_{dd}\\
A_{l+1} &= E_l\,P_{ed} + D_l\,(P_{de}+P_{dd}).
\end{split}
\label{eq:8}
\end{equation}
Combining Eqs.~(\ref{eq:7}) and (\ref{eq:8}), and using $P_{ed}=1$ [see Eqs.~(\ref{eq:2})], we obtain the following system of mutually recursive linear sequences,
\begin{equation}
\centering
\begin{pmatrix}
D_{l+1}\\
A_{l+1}
\end{pmatrix}
=
\begin{pmatrix}
(P_{dd} -1) & 1 \\
(P_{de}+P_{dd}-1) & 1
\end{pmatrix}
\begin{pmatrix}
D_{l}\\
A_{l}
\end{pmatrix}.
\label{eq:9}
\end{equation}
Then, the roots of the characteristic polynomial linked to the $2\times2$ matrix of system (\ref{eq:9}), 
\begin{equation}
\centering
\lambda_{1,2} = 
\frac{P_{dd} \pm \sqrt{P_{dd}^2 + 4\,P_{de}}}{2},
\label{eq:10}
\end{equation}
can be used to express the general solution of $D_{l}$ as,
\begin{equation}
\centering
D_{l} = a\,\lambda_1^{l-1} + b\,\lambda_2^{l-1},
\label{eq:11}
\end{equation}
where constants $a$ and $b$ can be found by using Eqs.~(\ref{eq:10}) and (\ref{eq:11}), and the expressions for $D_{l=1}$ and $D_{l=2}$, 
\begin{equation}
\centering
\begin{pmatrix}
a\\
b
\end{pmatrix}=
\begin{pmatrix}
\lambda_1^{0} & \lambda_2^{0}\\
\lambda_1^{1} & \lambda_2^{1}
\end{pmatrix}^{-1}
\begin{pmatrix}
q \\
q\,P_{dd}+ (1-q)\,P_{ed}
\end{pmatrix}.
\label{eq:12}
\end{equation}

In the light of the described above, we have formally found the exact solution for $P_n$. In the following set of equations we summarize our result,
\begin{equation}
\begin{split}
\centering
P_n &= D_{\frac{n+2}{3}}(1-p), \\
P_{n+1} &= D_{\frac{n+2}{3}}\,p\,(1-q)\,(1-p), \\
P_{n+2} &= D_{\frac{n+2}{3}}\,p\,(1-q)\,p\,(1-q)\,(1-p),
\end{split}
\label{eq:13}
\end{equation}
with $n=3l-2$, and $l\in\mathbb{N}$.

\begin{figure}[t!]
\centering
\includegraphics[width=0.8\textwidth]{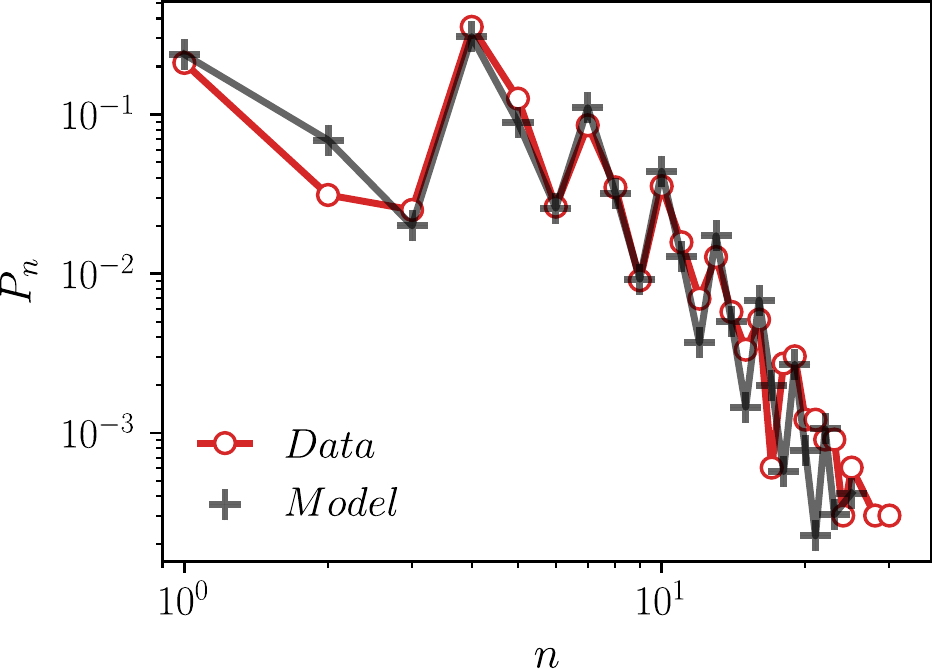}
\caption{
The probability that the players perform $n$ hits during rallies, $P_n$.
Colored in red with circles marks, the empirical case.
Colored in black with plus symbols, the analytical results.
We can observe that the analytical approach captures well the non-trivial behavior of the empirical distribution. The similarity can be quantified by calculating the Jensen-Shannon divergence between the curves. From this procedure, we obtain $D_{JS}=0.009$, which indicates a high similarity.
}
\label{fi:pn}
\end{figure}

In Fig.~\ref{fi:pn} we show the probability distribution $P_n$. 
The extracted from the data is colored in red and the theoretical calculated with equations (\ref{eq:13}) is colored in black. 
The values of parameters $p$ and $q$ were set by carrying out a minimization process using the Jensen-Shannon divergence ($D_{JS}$) between the curves as metrics.
From this procedure, we obtained $p=0.54$ and $q=0.46$ with $D_{JS}=0.009$, indicating a good similarity between the curves. Such correspondence can also be observed with the naked eye in the figure.
On the other hand, the mean value and the standard deviation for the empirical curve are $\avg{n}= 4.58(70)$, $\sigma(n)=3.40(45)$, and for the theoretical curve are $\avg{n^{TH}}= 4.74(83)$, $\sigma(n^{TH})=3.44(66)$.
We can see that the values are smaller in the empirical case but still very similar to the calculations.

Regarding the obtained values of $p$ and $q$, it is necessary to highlight that in a real games, each player on the court should have a different pair of values $p$ and $q$. This is because distinct players should perform differently.
Moreover, these pairs could depend on the set because players vary their performance during the game. 
However, with these results, we show that the simplification proposed in our model allows us to represent each player as a single {\it average player}. In this context, since we construct the empirical probability $P_n$ with the results of several matches, teams, and players, it is expected that throughout the minimization process, we obtain values close to $0.5$.

The reader might note that the shape of the curves follows a zig-zag pattern with the peaks placed at the values $n=1,4,7,10,...$, which are the number of hits related to teams' complete ball possessions.
We can explain this non-trivial behavior by analyzing the natural dynamics of the game. 
If a team can control an attack in the first hit, they will use the rest of the reglementary hits to set the ball and attack back. As we have previously mentioned, it is unlikely they attack at the first or second hit. 
Because of this, it is more probable to find in the dataset cases, for instance, with $n=7$ than cases with $n=6$.

We want to highlight that despite the complex dynamics of the game, the analytical approach based on our parsimonious stochastic model succeeds in capturing the non-trivial behavior of this relevant observable.

\section{Multi-agent simulations}
\begin{figure}[t!]
\centering
\includegraphics[width=1.0\textwidth]{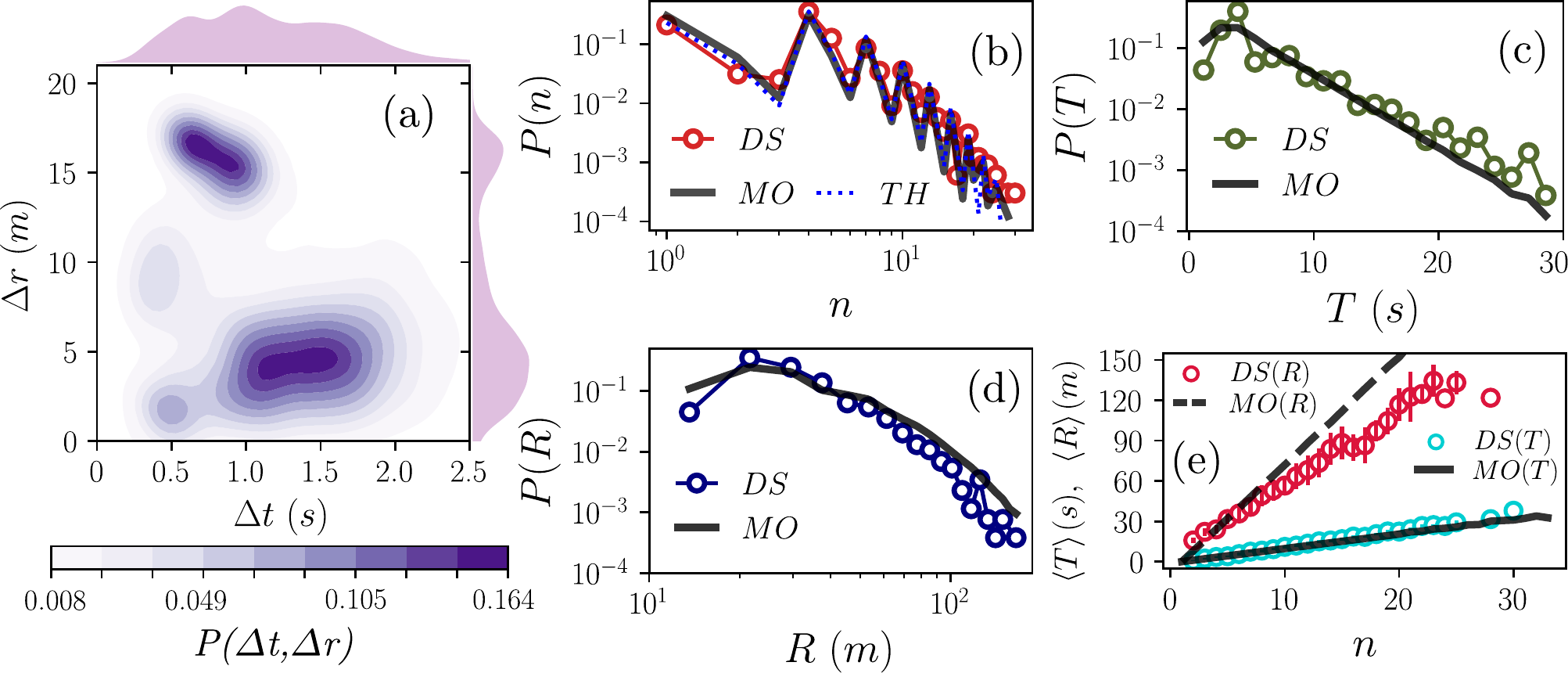}
\caption{
In these plots, we summarize the results obtained from the multi-agent simulations.
Panel (a) shows the joint probability distribution that we used to draw, at every simulation step, a pair $(\Delta t_i, \Delta r_i)$.
Panel (b) exhibits the probability distribution $P(n)$. Here we show the empirical observations (DS), the theoretical calculation (TH), and the obtained from the agent-based model (MO).
Panel (c) shows the probability distribution $P(T)$ obtained from the dataset (DS) and from the model's outcomes.
Panel (d) exhibits the probability distribution $P(R)$ obtained from the dataset (DS) and from the model's outcomes.
Panel (e) shows the relation $\avg{T}$ vs. $n$ and $\avg{R}$ vs. $n$, a comparison between the empirical case (DS) and the result of the model (MO).
}
\label{fi:simulations}
\end{figure}

We now apply the rules of our model to guide the dynamics of a minimalist 1-D self-propelled agents system thought of to emulate volleyball rallies.
We aim to confirm, with this agent-based model, the analytical results obtained for probability distribution $P_n$.
Moreover, we show that it is possible to combine the analytical results with the data to give an approach to capture other empirical global observables related to spatiotemporal variables.
The agent model is based on the following elements,
\begin{enumerate}
\item {\it Teams and players.} In this parsimonious system, there are two teams with two players (agents) by team. 

\item {\it The court.} We represent the game's court as a 1-D array with four sites. 
The 1st and the 2nd sites shape the first team side, and the 3rd and the 4th sites shape the second team side.
The net is placed between sites 2 and 3, delimiting the teams' sides.
Players are allowed to occupy any site of their zones. They can also overlap, but they cannot invade the rival's side.

\item {\it The teammates' roles.} To emulate the player's tactical dependency in volleyball teams, we propose that one of the two players manages the 1st and 3rd hit (defender/ attacker), and the other manages the 2nd hit (setter).

\item {\it Initial conditions}. The player's sites at the beginning of the rally are randomly set. The player at the 1st site serves the ball.

\item {\it Dynamics.} If the agents have to change their current place to hit the ball, then the parameters $p$ and $q$ give the probability of performing the hit and achieving precision, respectively (see section Model). 
Achieving precision in this context means, in the 1st hit, sending the ball to the partner place, and in the 3rd hit, sending the ball to the site where the rival in charge of taking the 1st hit is not occupying.
The rally ends when one of the players misses a hit.
\end{enumerate}
We define $T$  as the rallies' total time and $R$ as the length of the projection of the ball trajectory on the court's floor. 
These two variables can be empirically measured from the collected data by simply adding the succession of temporal and spatial intervals, $\Delta t$ and $\Delta r$, observed between rallies' events. 
To obtain $T$ and $R$ from simulations we propose the following,
\begin{enumerate}
\setcounter{enumi}{5}
\item {\it Global observables $T$ and $R$}. At every step of the dynamics, the players hitting the ball draws from the empirical joint distribution P($\Delta t$,$\Delta r$), shown in Fig.~\ref{fi:simulations}~(a), a pair of values (${\Delta t}_i$, ${\Delta r}_i$) that we will link to the $ith$ event. 
Then, at the end of the rallies, we compute $T= \sum^n_i {\Delta t}_i$ and $R= \sum^n_i {\Delta r}_i$, where $n$ is the total number of events, to obtain the simulated values of these observables.
\end{enumerate}
Notice we sample $\Delta t$ and $\Delta r$ from the joint distribution because these variables are correlated. 
This is evidenced in Fig.~\ref{fi:simulations}~(a), where we can see that the multimodal behavior of $P(\Delta r)$ and $P(\Delta t)$ results in a non-trivial joint distribution with several local maxima.

With the model defined, we performed $10^5$ rally simulations setting the parameters $p$ and $q$ with the values calculated in section \ref{se:analytical} .
In Fig.~\ref{fi:simulations} panels (b) to (e), we compare the outcomes with the empirical data. Additionally, in panel (b), we plot the theoretical curve given by the Eq.~(13). In the following, we discuss these results.
In panel (b), we show the probability distribution $P(n)$. We can see that it agrees with the theoretical curve, confirming the calculations.
Panel (c) show the probability distribution $P(T)$. 
In this case, we can see that the model captures very well the empirical data.
The plot is in linear-log scale to emphasize the probability decay as an exponential curve.
The calculation for the mean value and the standard deviation for the empirical data gives $\avg{T^{DS}}=5.47~(s)$ and $\sigma(T^{DS})=4.00~(s)$, and for the simulations $\avg{T^{MO}}=3.52~(s)$ and $\sigma(T^{MO})=3.86~(s)$.
The similarity between the values of the first and second moment supports the existence of an exponential behavior.
In panel (d), the probability distribution $P(R)$ is shown. As well as in previous cases, we see that the model satisfactorily captures the global aspects of the empirical data.
The mean and standard deviation for both cases are  
$\avg{R^{DS}}=34.42~(m)$, $\sigma(R^{DS})=18.23~(m)$, and  $\avg{R^{MO}}=39.17~(m)$, $\sigma(R^{MO})=25.52~(m)$.
Lastly, in panel (e), we show the evolution of $\avg{T}$ and $\avg{R}$ with the number of hits $n$.
We can see a linear behavior in both cases. 
For $\avg{T}$ the empirical observations agree well with the model's outcomes. 
For $\avg{R}$, on the contrary, we observe a deviation at higher values of $n$.
This deviation may be related to the emergence of some type of complexity in the actual dynamics that our simple model cannot capture.
For instance, sometimes during the game the play becomes unstable, and the players hit the ball several times in reduced space until one team is able to control the ball.
Notice we cannot capture this kind of ``burstiness effect'' through our simple model.
According to the stated rules, at every step, we have to randomly draw a pair $(\Delta t, \Delta r)$ which implies following a memoryless process.
In this frame, since we have a joint probability distribution that is bimodal in the variable $\Delta r$,  the probability of obtaining a long sequence of pairs with small values of $\Delta r$ is low. 
Consequently, for a given value of $n$, the calculated value of $\avg{R}$ in the model could tend to be slightly higher than in the empirical case, as we observe in Fig.~\ref{fi:simulations} panel (e).

\section{Conclusions}

Our investigation focused on studying the dynamics of volleyball's rallies. 
We proposed a framework based on games visualization,  collection of relevant information, and through data-driven analysis aiming to obtain insights to define the rules of a mathematical model able to capture the underlying dynamics in volleyball games.
We found that the players are more likely to fail and become imprecise if they have to move out of their actions zones to perform a hit. 
With this in mind, we proposed a model based on two stochastic parameters: $p$ and $q$, where the first is the probability of performing the hit, and the latter is the probability of achieving precision. 
Then we calculated a closed-form expression for the probability that players perform $n$ hits in the rally, $P_n$, that agree remarkably well with the empirical observations.
Therefore, we understand that we have uncovered two stochastic variables able to partially generate the level of complexity observed in the volleyball dynamics. 
In this regard, we consider that this work represents a new step towards a broad understanding of volleyball games as complex adaptive systems.
Moreover, the collected data that we make publicly available with this work is, as far as we know, the largest open collection of volleyball-logs ever released, an invaluable resource for the research community that opens the door for further research in this area.
Finally, we want to point out that our findings provide new knowledge that should be actively taken into account for sports scientists and coaches.
Since it has been shown that the length of rallies may affects the team behavior through many indirect variables \cite{sanchez2015analysis,link2019performance,sanchez2016dynamics,mroczek2014analysis}, our results can be handy for designing new efficient data-driven training systems aiming to enhance the performance in competitive scenarios.

\section*{Data availability statement}

The data that support the findings of this study are publicly available in \cite{data}.

\section*{Acknowledgement}

We acknowledge enlightening discussions with Luc\'ia Pedraza.
This work was partially supported by CONICET under Grant number PIP 112 20200 101100; FonCyT under Grant number PICT-2017-0973; and SeCyT-UNC (Argentina).


\begin{thebibliography}{10}

\bibitem{blasius2009zipf}
Bernd Blasius and Ralf T{\"o}njes.
\newblock Zipf’s law in the popularity distribution of chess openings.
\newblock {\em Physical Review Letters}, 103(21):218701, 2009.

\bibitem{clauset2015safe}
A~Clauset, M~Kogan, and S~Redner.
\newblock Safe leads and lead changes in competitive team sports.
\newblock {\em Physical Review E}, 91(6):062815, 2015.

\bibitem{holleczek2012particle}
Thomas Holleczek and Gerhard Tr{\"o}ster.
\newblock Particle-based model for skiing traffic.
\newblock {\em Physical Review E}, 85(5):056101, 2012.

\bibitem{baek2015nash}
Seung~Ki Baek, Seung-Woo Son, and Hyeong-Chai Jeong.
\newblock Nash equilibrium and evolutionary dynamics in semifinalists' dilemma.
\newblock {\em Physical Review E}, 91(4):042144, 2015.

\bibitem{ben2007efficiency}
E~Ben-Naim and NW~Hengartner.
\newblock Efficiency of competitions.
\newblock {\em Physical Review E}, 76(2):026106, 2007.

\bibitem{yamamoto2021preferential}
Ken Yamamoto and Takuma Narizuka.
\newblock Preferential model for the evolution of pass networks in ball sports.
\newblock {\em Physical Review E}, 103(3):032302, 2021.

\bibitem{neiman2011reinforcement}
Tal Neiman and Yonatan Loewenstein.
\newblock Reinforcement learning in professional basketball players.
\newblock {\em Nature communications}, 2(569):1--8, 2011.

\bibitem{mukherjee2019prior}
Satyam Mukherjee, Yun Huang, Julia Neidhardt, Brian Uzzi, and Noshir
  Contractor.
\newblock Prior shared success predicts victory in team competitions.
\newblock {\em Nature human behaviour}, 3(1):74--81, 2019.

\bibitem{mandic2019trends}
Radivoj Mandi{\'c}, Sa{\v{s}}a Jakovljevi{\'c}, Frane Er{\v{c}}ulj, and Erik
  {\v{S}}trumbelj.
\newblock Trends in nba and euroleague basketball: Analysis and comparison of
  statistical data from 2000 to 2017.
\newblock {\em PloS one}, 14(10), 2019.

\bibitem{merritt2013environmental}
Sears Merritt and Aaron Clauset.
\newblock Environmental structure and competitive scoring advantages in team
  competitions.
\newblock {\em Scientific reports}, 3:3067, 2013.

\bibitem{gudmundsson2017spatio}
Joachim Gudmundsson and Michael Horton.
\newblock Spatio-temporal analysis of team sports.
\newblock {\em ACM Computing Surveys (CSUR)}, 50(2):1--34, 2017.

\bibitem{fister2015computational}
Iztok Fister~Jr, Karin Ljubi{\v{c}}, Ponnuthurai~Nagaratnam Suganthan,
  Matja{\v{z}} Perc, and Iztok Fister.
\newblock Computational intelligence in sports: challenges and opportunities
  within a new research domain.
\newblock {\em Applied Mathematics and Computation}, 262:178--186, 2015.

\bibitem{patel2020intertwine}
Devansh Patel, Dhwanil Shah, and Manan Shah.
\newblock The intertwine of brain and body: a quantitative analysis on how big
  data influences the system of sports.
\newblock {\em Annals of Data Science}, 7(1):1--16, 2020.

\bibitem{drikos2009correlates}
Sotiris Drikos, Panagiotis Kountouris, Alexandros Laios, and Yiannis Laios.
\newblock Correlates of team performance in volleyball.
\newblock {\em International Journal of Performance Analysis in Sport},
  9(2):149--156, 2009.

\bibitem{buldu2019defining}
Javier~M Buldu, J~Busquets, Ignacio Echegoyen, et~al.
\newblock Defining a historic football team: Using network science to analyze
  guardiola’s fc barcelona.
\newblock {\em Scientific reports}, 9(13602):1--14, 2019.

\bibitem{bransen2019measuring}
Lotte Bransen, Jan Van~Haaren, and Michel van~de Velden.
\newblock Measuring soccer players’ contributions to chance creation by
  valuing their passes.
\newblock {\em Journal of Quantitative Analysis in Sports}, 15(2):97--116,
  2019.

\bibitem{cakmak2018computational}
Ali Cakmak, Ali Uzun, and Emrullah Delibas.
\newblock Computational modeling of pass effectiveness in soccer.
\newblock {\em Advances in Complex Systems}, 21(03n04):1850010, 2018.

\bibitem{gama2016networks}
Jos{\'e} Gama, Gon{\c{c}}alo Dias, Micael Couceiro, Tiago Sousa, and Vasco Vaz.
\newblock Networks metrics and ball possession in professional football.
\newblock {\em Complexity}, 21(S2):342--354, 2016.

\bibitem{garrido2020consistency}
D~Garrido, DR~Antequera, J~Busquets, R~L{\'o}pez Del~Campo, R~Resta Serra,
  S~Jos Vielcazat, and JM~Buld{\'u}.
\newblock Consistency and identifiability of football teams: a network science
  perspective.
\newblock {\em Scientific Reports}, 10(19735):1--10, 2020.

\bibitem{jusup2022social}
Marko Jusup, Petter Holme, Kiyoshi Kanazawa, Misako Takayasu, Ivan Romi{\'c},
  Zhen Wang, Sun{\v{c}}ana Ge{\v{c}}ek, Tomislav Lipi{\'c}, Boris Podobnik, Lin
  Wang, et~al.
\newblock Social physics.
\newblock {\em Physics Reports}, 948:1--148, 2022.

\bibitem{chacoma2020modeling}
A~Chacoma, N~Almeira, JI~Perotti, and OV~Billoni.
\newblock Modeling ball possession dynamics in the game of football.
\newblock {\em Physical Review E}, 102(4):042120, 2020.

\bibitem{chacoma2021stochastic}
A~Chacoma, N~Almeira, JI~Perotti, and OV~Billoni.
\newblock Stochastic model for football's collective dynamics.
\newblock {\em Physical Review E}, 104(2):024110, 2021.

\bibitem{sanchez2015analysis}
J~S{\'a}nchez-Moreno, R~Marcelino, I~Mesquita, and A~Ure{\~n}a.
\newblock Analysis of the rally length as a critical incident of the game in
  elite male volleyball.
\newblock {\em International Journal of Performance Analysis in Sport},
  15(2):620--631, 2015.

\bibitem{link2019performance}
Daniel Link and Sebastian Wenninger.
\newblock Performance streaks in elite beach volleyball-does failure in one
  sideout affect attacking in the next?
\newblock {\em Frontiers in psychology}, 10:919, 2019.

\bibitem{sanchez2016dynamics}
Joaqu{\'\i}n S{\'a}nchez-Moreno, Jos{\'e} Afonso, Isabel Mesquita, and Aurelio
  Ure{\~n}a.
\newblock Dynamics between playing activities and rest time in high-level
  men’s volleyball.
\newblock {\em International Journal of Performance Analysis in Sport},
  16(1):317--331, 2016.

\bibitem{mroczek2014analysis}
Dariusz Mroczek, Aleksander Januszkiewicz, Adam~S KawczyNski, Zbigniew
  Borysiuk, and Jan Chmura.
\newblock Analysis of male volleyball players' motor activities during a top
  level match.
\newblock {\em The Journal of Strength \& Conditioning Research},
  28(8):2297--2305, 2014.

\bibitem{data}
A.~Chacoma.
\newblock Github repository.
\newblock \url{https://github.com/chacoma/volleyball}, 2022.

\end{thebibliography}
\end{document}